\documentclass[12pt]{article}

\usepackage{scicite}

\usepackage{times}

\usepackage{graphicx}
\usepackage{color}

\newenvironment{sciabstract}{%
\begin{quote} \bf}
{\end{quote}}

\newcounter{lastnote}
\newenvironment{scilastnote}{%
\setcounter{lastnote}{\value{enumiv}}%
\addtocounter{lastnote}{+1}%
\begin{list}%
{\arabic{lastnote}.}
{\setlength{\leftmargin}{.22in}}
{\setlength{\labelsep}{.5em}}}
{\end{list}}

\title{ Universality of market superstatistics }

\author
{Mateusz Denys$^{1}$, Maciej Jagielski$^{\ast 1,2,3}$, Tomasz
  Gubiec$^{1}$, Ryszard Kutner$^{1}$, and \\  
H. Eugene Stanley$^{2}$ \\
\\
\normalsize{$^{1}$Faculty of Physics, University of Warsaw, Pasteura Str. 5, 
PL-02093 Warsaw, Poland}\\ 
\normalsize{$^{2}$Center for Polymer Studies and Dept. of Physics,
  Boston Univ., Boston, MA 02215 USA}\\ 
  \normalsize{$^{3}$Department of Management, Technology and Economics, ETH Z\"urich, }\\
  \normalsize{$^{}$ Scheuchzerstrasse 7, 
  CH-8092 Z\"urich, Switzerland}\\
\\
\normalsize{$^\ast$To whom correspondence should be addressed; E-mail:
  zagielski@gmail.com.} 
}

\date{{}}

\begin{document} 


\baselineskip14pt


\maketitle 

\begin{sciabstract}

We use a continuous-time random walk (CTRW) to model market fluctuation
data from times when traders experience excessive losses or excessive
profits. We analytically derive ``superstatistics'' that accurately
model empirical market activity data (supplied by Bogachev, Ludescher,
Tsallis, and Bunde) that exhibit transition thresholds. We measure the
interevent times between excessive losses and excessive profits, and
use the mean interevent time as a control variable to derive a
universal description of empirical data collapse.  Our superstatistic
value is a weighted sum of two components, (i) a power-law corrected by
the lower incomplete gamma function, which asymptotically tends toward
robustness but initially gives an exponential, and (ii) a power-law
damped by the upper incomplete gamma function, which tends toward the power-law
only during short interevent times.  We find that the scaling shape
exponents that drive both components subordinate themselves and a
``superscaling'' configuration emerges. We use superstatistics to
describe the hierarchical activity when component (i) reproduces the
negative feedback and component (ii) reproduces the stylized fact of
volatility clustering. Our results indicate that there is a functional
(but not literal) balance between excessive profits and excessive losses
that can be described using the same body of superstatistics, but
different calibration values and driving parameters.

\end{sciabstract}

\section{Introduction}

Financial markets fluctuate as traders estimate risk levels and strive
to make a profit. The interevent interval between times when market
returns are producing excessive profits and times when they are
producing excessive losses can be described using a continuous-time
random walk (CTRW) formalism (see Refs.~\cite{GPHS,JHKK,RKFS,SChKM} and
references therein).

Empirical market data on excessive profits and losses
\cite{LTB,LB,BB1,BB2} define excessive profits as those greater than
some positive fixed threshold $Q$ and excessive losses as those below
some negative threshold $-Q$. The mean interevent time\footnote{The term
  `interevent time' appears in the literature under such names as
  `pausing time', `waiting time', `intertransaction time' and
  `interoccurrence time' in the context of different versions of the
  continuous-time random walk formalism \cite{KKB,PMKK,KKPM,GK,SChKM}.}
between losses versus $Q$ has been used as an aggregated basic variable.

Interevent times constitute a universal stochastic measurement of
market activity on time\-scales that range from one minute to one month
\cite{LTB,LB}. The mean interevent time can be used as a control
variable that produces a universal description of empirical data
collapse \cite{BB1}, i.e., the distribution of interevent times for a
fixed mean interevent time is a universal statistical quantity
unaffected by time scale, type of market, asset, or index.

This distribution can be described using (i) the CTRW valley model (see
Refs.~\cite{SChKM,JHKK} and references therein), which treats time
intervals as random variables, and (ii) generalized extreme value
statistics\footnote{Whether the value of losses or profits in the basic
  stochastic process are statistically independent is irrelevant because
  any possible correlations between them are absent in our derivations.}
for stochastic dependent basic processes \cite{BC}, which are
$q$-exponentials not \emph{ad hoc} statistics \cite{LTB,LB}.
Inter-event times in a multifractal structure of financial markets
\cite{PMKK,KKPM} and in the single-step memory in order-book transaction
dynamics \cite{GK} are foundational in the analysis of double-action
market activity.

\section{Principal goal}

Our principal goal is to model the empirical data\footnote{All data fits
  and drawings were made using Mathematica Ver. 10.} associated with
single-variable statistics, i.e., (i) the mean inter-event time period
$R_Q$ between extreme (excessive) losses, defined as those below a
negative threshold $-Q$, as a function of the $Q(>0)$ value\footnote{For
  the sake of simplicity, we will treat losses as positive quantities.}
and (ii) the distribution $\psi _Q(\Delta _Q t)$ of inter-event times
between losses $\Delta _Q t$, previously described using {\it ad hoc}
$q$-exponentials.

Because no empirical data associated with item (i) are available, in our
study of excessive profits we will focus on item (ii) and use the
empirical data provided in Refs.~\cite{LTB,LB,BB1,BB2}. Note that the
$q$-exponentials used in Refs.~\cite{LTB,LB,BB1,BB2} cannot produce the
key empirical data in item (i), and thus in our approach we use
superstatistics. Because small losses and profits are of little concerm
to traders, we focus on medium to high $Q$-values. Our goal is to
provide market superstatistics that have universality.

\section{Basic achievement}\label{section:basach}

We here find an analytical closed form of the mean interevent time
period $R_Q$ between excessive (extreme) losses that is greater than
some threshold $Q$, i.e.,
\begin{eqnarray}
\tau R_Q^{-1}=P(-\varepsilon \leq -Q)=P(\varepsilon \geq
Q)=\int_{Q}^{\infty}D(\varepsilon)d\varepsilon , 
\label{rown:RQ-1}
\end{eqnarray}
where $\tau $ is a time unit\footnote{Later in this text we will set $\tau
  =1$.} and $D(\varepsilon)$ the density of returns given by the Weibull
distribution of extreme (or excessive) losses \cite{DS,EKM,FHH},
\begin{eqnarray}
D(\varepsilon )=\frac{\eta }{\bar{\varepsilon }}\left(\frac{\varepsilon
}{\bar{\varepsilon }}\right)^{\eta -1} 
\exp\left(-\left(\frac{\varepsilon }{\bar{\varepsilon }}\right)^{\eta
}\right), \; \bar{\varepsilon },\; \eta >0. 
\label{rown:rho}
\end{eqnarray}

Note that we consider random variable $\varepsilon$ to be an increment
of some underlying stochastic process.\footnote{For the Weibull
  distribution the relative mean value $\frac{\langle \varepsilon
    \rangle }{\bar{\varepsilon }}=\frac{1}{\eta }\, \Gamma (1/\eta )$
  and the relative variance $\frac{\sigma ^2}{\langle \varepsilon
    \rangle ^2}=\frac{\langle \varepsilon ^2\rangle -\langle \varepsilon
    \rangle ^2}{\langle \varepsilon \rangle ^2}=\left[2\eta \frac{\Gamma
      (2/\eta )}{\Gamma ^2(1/\eta )}-1\right]$ are $\eta $-dependent
  that is, they are (for fixed exponent $\eta $) universal quantities.}
Values of this random variable can, in general, be dependent
\cite{BC}. Here we consider case $\eta <1$ (see Table~\ref{table:tab1})
which means that distribution $D(\varepsilon )$ is, for $\varepsilon
/\bar{\varepsilon }\gg 1$, a decreasing truncated power-law \cite{BBM}.

Reference~\cite{IYPL} uses the Weibull distribution to describe the
statistics of interevent times between subsequent transactions for a
given asset.  We use the Weibull distribution and the conditional
exponential distribution of the CTRW valley model to derive
superstatistics or complex statistics associated with the threshold of
excessive losses or profits.
 
Substituting (\ref{rown:rho}) into (\ref{rown:RQ-1}), we obtain
\begin{eqnarray}
R_Q=\exp\left( \left( \frac{Q}{\bar{\varepsilon }}\right)^{\eta }\right),
\label{rown:RQ}
\end{eqnarray}
i.e., $\ln R_Q$ increases vs. the relative variable $Q/\bar{\varepsilon
}$ according to a power-law.

\begin{figure}
\begin{center}
\includegraphics[width=145mm,angle=0,clip]{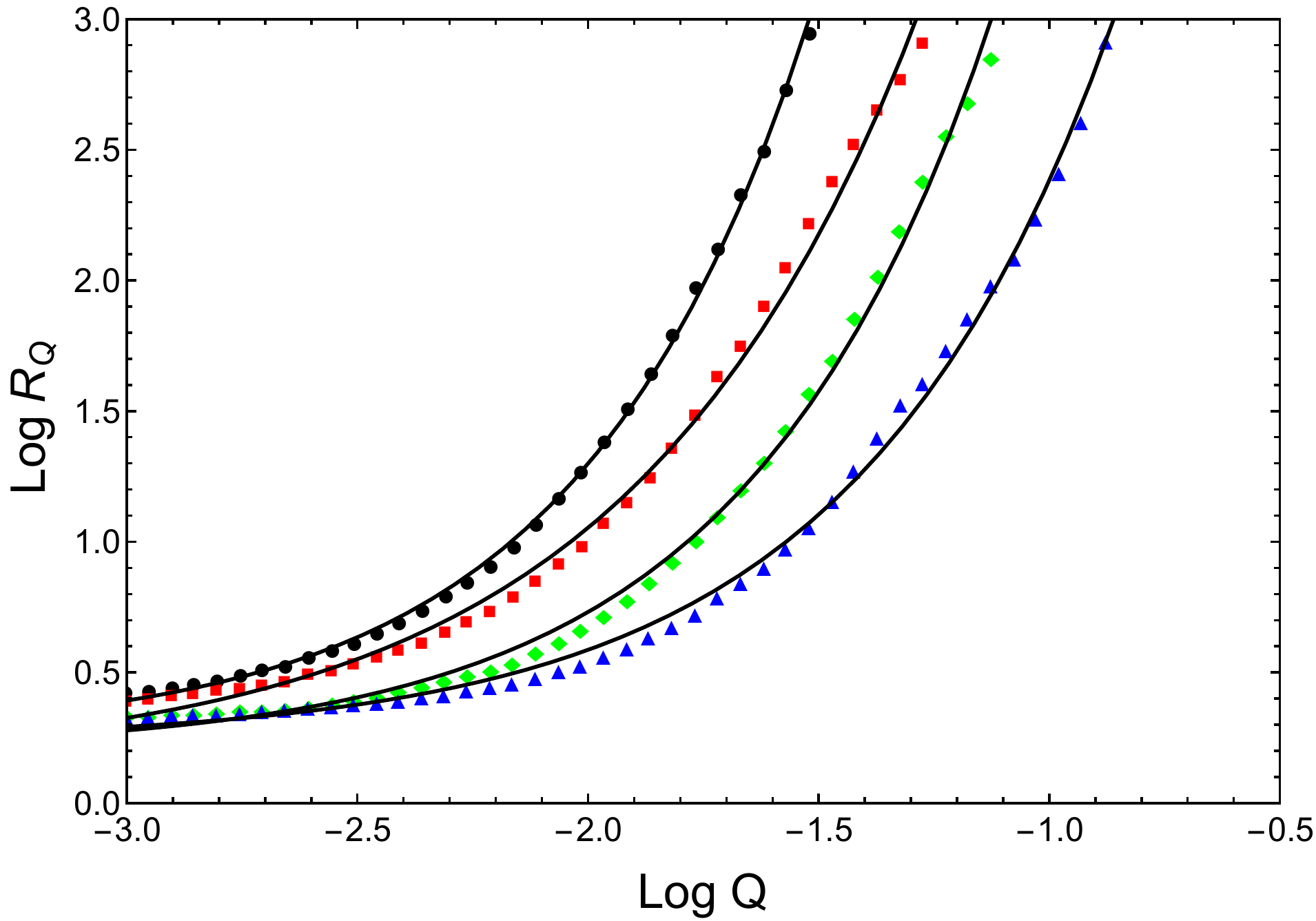}
\caption{Mean interevent time period $R_Q$ vs. threshold $Q$ for four
  typical classes of quotations. Black circles, red squares, green
  rhomboids, and blue triangles concern US/GBP exchange rate, S\&P 500
  index, IBM stock, and WTI (the crude oil) empirical data (from January
  2000 to June 2010), respectively, taken from Fig. 2 in ref. \cite{LTB}
  (plotted from the top curve down to the bottom one.) The solid, well
  fitted curves present predictions of our formula (\ref{rown:RQ}) -
  values of their fitted parameters $\bar{\varepsilon }$ and $\eta $ are
  shown in tab. \ref{table:tab1} (the unimportant multiplicative
  calibration parameter is not presented there). Subtle wavy deviations
  from these predictions are not considered in this work. (Empirical
  data were used by permission of the EPL.)}
\label{figure:fig.2}
\end{center}
\end{figure}

The solid curves in Fig.~\ref{figure:fig.2} indicate the predictions
generated by (\ref{rown:RQ})\footnote{Note that each curve has a
  slightly different multiplicative calibration parameter that defines
  its vertical shift.} and fit the empirical data (the points are
represented by different marks). This basic agreement enables us to
construct the corresponding superstatistics and allows us to study the
successive empirical data (given in Secs.~\ref{section:superstat} and
\ref{section:disconc}). Because the statistical error is low we are able
to determine $\eta$ and $\bar{\varepsilon}$ and derive the subsequent
parameters that define the shape of the superstatistics. For example,
the empirical data in Fig.~\ref{figure:fig.2} indicates that the value
of the shape parameter exponent is $\eta<1$
(cf. Table~\ref{table:tab1}), which for large losses or profits (i.e.,
$\varepsilon \gg \bar{\varepsilon }$) makes the Weibull distribution
(\ref{rown:rho}) a stretched exponentially truncated power-law. In
contrast, the results presented in Ref.~\cite{LTB} are incomplete
because they allow no analogous comparison with theoretical predictions
based on the $q$-exponential.

\begin{table}
\centering
\caption{Values of exponent $\eta $ and quantity $\bar{\varepsilon }$
  obtained from the fit of predictions of formula (\ref{rown:RQ}) to the
  empirical data (all of them plotted in Fig. \ref{figure:fig.2})
  representing the exchange rate US Dollar against Great British Pound,
  the index S\&P 500, the IBM stock, and crude oil (WTI).}
 \begin{tabular}{|c||c|c|}
  \hline 
 Index/Par. & $\eta $ & $\bar{\varepsilon}$ \\ 
\hline  \hline
  US/GBP & 0.8756$\pm $0.0156 & 0.0037$\pm $0.0003 \\
  \hline 
	S\&P500 & 0.6981$\pm $0.0292 & 0.0035$\pm $0.0005 \\
  \hline
	IBM & 0.8246$\pm $0.0236 & 0.0078$\pm $0.0007 \\
	\hline
	WTI & 0.7855$\pm $0.0182 & 0.0131$\pm $0.0008 \\
  \hline
\end{tabular} 
\label{table:tab1}
\end{table}

\section{Superstatistics}\label{section:superstat}

We next construct an unnormalized, unconditional distribution $\psi
_Q(\Delta _Q t)$ of the interevent time stochastic variable, $\Delta _Q
t$, in the form of superstatistics\footnote{To normalize the
  superstatistics given by (\ref{rown:psiDt}) we divide $\psi _Q(\Delta
  _Q t)$ by $\int_Q^{\infty}D(\varepsilon )d\varepsilon
  =\exp\left(-\left( \frac{Q}{\bar{\varepsilon }}\right)^{\eta }\right)$
  or multiply it by $R_Q$. This produces conditional superstatistics
  limited to profits and losses no smaller then $Q$.}  based on the
Weibull distribution used in Sec.~\ref{section:basach},
\begin{eqnarray}
\psi _Q(\Delta _Qt)=\int_Q^{\infty }\psi _Q(\Delta _Q t|\varepsilon
)D(\varepsilon )d\varepsilon.
\label{rown:psiDt}
\end{eqnarray}
Here we assume the conditional distribution $\psi _Q(\Delta _Q
t|\varepsilon )$ is in the exponential form\footnote{The exponential
  form of the conditional distribution (\ref{rown:psiDte}) assumes that
  the losses or profits of a {\it fixed\/} value $\varepsilon $ are
  statistically independent, which is generally not valid for different
  values of losses and profits.}
\begin{eqnarray}
\psi _Q(\Delta _Q t|\varepsilon )=\frac{1}{\tau _Q(\varepsilon
  )}\exp\left(-\frac{\Delta _Q t}{\tau _Q(\varepsilon )}\right).
\label{rown:psiDte}
\end{eqnarray}
Because it is conditional, the next (subsequent) loss is exactly
$\varepsilon$, and the relaxation time is given by the stretched
exponential
\begin{eqnarray}
\tau _Q(\varepsilon )=\tau _Q(0)\exp\left((B_Q\varepsilon )^{\eta }\right)
\label{rown:taue}
\end{eqnarray}
as a straightforward extension of the exponential relaxation time used
in the canonical version of the CTRW introduced by
\cite{MW,GHW,GPHS,HSM} in the context of photocurrent relaxation in
amorphous films. Here $\tau _Q(0)$ is a free ($\varepsilon$-independent)
relaxation time, and quantity $B_Q(>0)$ is independent of variable
$\varepsilon$.  Quantity $B_Q$ is a formal analog of an inverse
temperature and we will later derive its scaling with the control
threshold $Q$ value. We use the $\eta$ exponent in (\ref{rown:rho}) to
reduce the number of free exponents in (\ref{rown:taue}) (Ockham's razor
principle) and to derive superstatistics $\psi _Q(\Delta _Qt)$ in an
exact closed analytical form. Note that the stochastic dependence of
interevent time $\Delta_Qt$ on loss $\varepsilon $ assumed in
(\ref{rown:psiDte}) is confirmed when smaller losses appear more
frequently than larger ones. This is described by definition
(\ref{rown:taue}) in which mean time $\langle \Delta_Qt\rangle
_{\varepsilon }=\tau _Q(\varepsilon )$ is a monotonically increasing
function of $\varepsilon $. This creates an expanding hierarchy of
interevent times where larger losses and profits appear {\it less\/}
frequently than smaller ones. To make larger losses or profits appear
{\it more\/} frequently than smaller ones, we create the opposite
hierarchy of losses and profits using
\begin{eqnarray}
\tau _Q^{\prime }(\varepsilon )=\tau _Q(0) \exp\left(-(B_Q\varepsilon
)^{\eta }\right). 
\label{rown:tauen}
\end{eqnarray}
In this opposite case (which is also analytically solvable) we encounter
a clustering phenomenon in which shorter time intervals separate the
larger values of losses/profits rather than the smaller ones. This
complementary case is briefly discussed in Sec.~\ref{section:disconc}.

Substituting (\ref{rown:taue}) and (\ref{rown:psiDte}) into
(\ref{rown:psiDt}) we finally derive a superstatistics in the searched
form
\begin{eqnarray}
\psi _Q(\Delta _Q t)&=&\frac{1}{\tau _Q(Q)}\frac{\alpha _Q}{(\Delta _Q
  t/\tau _Q(Q))^{1+\alpha _Q}} \nonumber \\ 
&\times &{\bf \gamma }_{Euler}(1+\alpha _Q,\Delta _Q t/\tau _Q(Q)),
\label{rown:psiDtfin}
\end{eqnarray}
where the scaling shape exponent 
\begin{eqnarray}
\alpha _Q=\frac{1}{(B_Q \bar{\varepsilon })^{\eta }}=\frac{1}{\ln
  \left(\tau _Q(\bar{\varepsilon })/\tau _Q(0)\right)}, 
\label{rown:alphaQ}
\end{eqnarray} 
and the lower incomplete gamma function 
\begin{eqnarray}
{\bf \gamma }_{Euler}(1+\alpha _Q,\Delta _Q t/\tau _Q(Q))=\int_0^{\Delta
  _Q t/\tau _Q(Q)} y^{\alpha _Q}\exp(-y)dy. 
\label{rown:gameul}
\end{eqnarray}

The significant step in the derivation of formula (\ref{rown:psiDtfin})
is the replacement of the running variable $\varepsilon$, present in the
integration (\ref{rown:psiDt}), with a new running variable $y=\Delta
_Qt/\tau _Q(\varepsilon )$. This changes the stretched exponential to
exponential in the overall function under the integral in
(\ref{rown:psiDt}) making the integration a straightforward (exact and
closed) operation. Note that the first equality in (\ref{rown:alphaQ})
gives a straightforward, formal generalization of the corresponding
exponent obtained within the canonical CTRW valley model
\cite{MW,GHW,GPHS,HSM}, where $B_Q$ is the thermodynamic $\beta$,
$\bar{\varepsilon}$ is the mean valley depth, and the exponent value
is $\eta=1$.

Equation (\ref{rown:psiDtfin}) asymptotically (for $\Delta _Q t/\tau
_Q(Q)\gg 1$) takes a power-law form
\begin{eqnarray}
\psi _Q(\Delta _Q t)\approx \frac{1}{\tau _Q(Q)}\frac{\alpha _Q}{(\Delta
  _Q t/\tau _Q(Q))^{1+\alpha _Q}}{\bf \Gamma }_{Euler}(1+\alpha _Q)
\label{rown:psiDtasympt}
\end{eqnarray}
of the relative interevent time $\Delta _Q t/\tau _Q(Q)$ while initially
(for $\Delta _Q t/\tau _Q(Q)\ll 1$) it takes an exponential form
\begin{eqnarray}
\psi _Q(\Delta _Q t)\approx \frac{1}{\tau _Q(Q)}\frac{\alpha
  _Q}{1+\alpha _Q}\exp\left(-\frac{1+\alpha _Q}{2+\alpha _Q}\Delta _Q
t/\tau _Q(Q)\right).
\label{rown:psiDtinit}
\end{eqnarray}

For $\alpha _Q\gg 1$, Eq.~(\ref{rown:psiDtfin}) reduces to the $\alpha
_Q$-independent exponential
\begin{eqnarray}
\psi _Q(\Delta _Q t)\approx \frac{1}{\tau _Q(Q)}\exp\left(-\Delta _Q t/\tau _Q(Q)\right),
\label{rown:psiDtinitexp}
\end{eqnarray}
which is consistent with Eq.~(\ref{rown:psiDtinit}).

\begin{figure}
\begin{center}
\includegraphics[width=150mm,angle=0,clip]{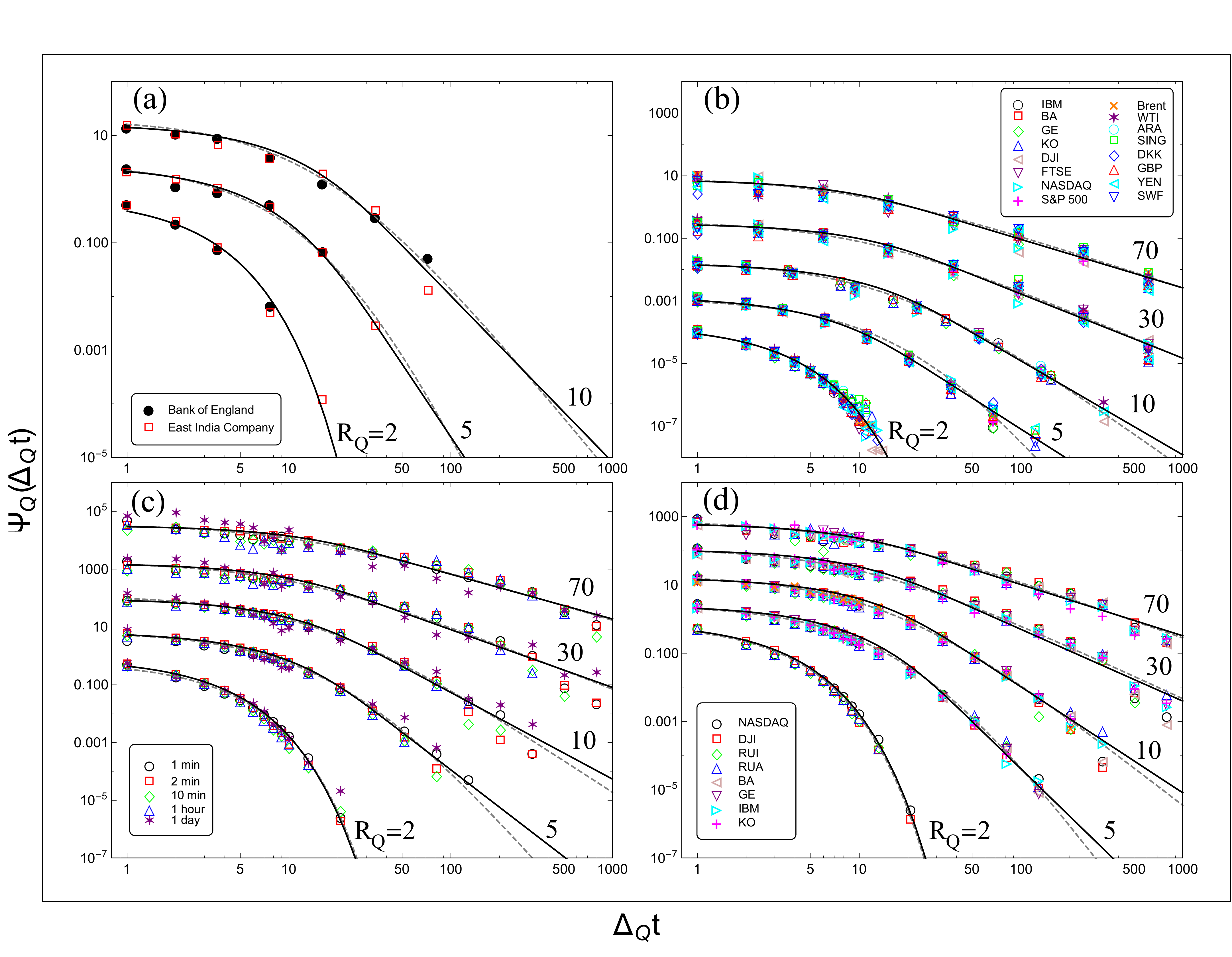}
\caption{Collected plots of empirical distributions (colored marks) and
  theoretical superstatistics, $\psi _Q(\Delta _Qt)$, (black solid
  curves), which are predictions of our formula (\ref{rown:psiDtfin})
  (while the dashed curves were given by q-exponential shown by eq. (3)
  in \cite{LTB}) vs. interevent time, $\Delta _Qt$, for the monthly
  returns (a), for the relative daily price returns for sixteen typical
  examples of financial data in the period 1962-2010 (b), from minutes
  over the hours to daily returns for NASDAQ between March 16, 2004 and
  June, 2006 (c), and for the detrended minute-by-minute eight most
  typical examples of financial data (d). (All empirical data were drawn
  from \cite{LTB,LB} for permission of EPL and PRE.)}
\label{figure:Fig_3_7_8_9}
\end{center}
\end{figure}

Note that our approach is based solely on the relaxation time
(\ref{rown:taue}) as a function of a single variable $\varepsilon$.
Only $\varepsilon =0, \bar{\varepsilon }, Q$, is used, and parameter $Q$
is the external control threshold, i.e., by using (\ref{rown:taue}) and
(\ref{rown:RQ}), we obtain
\begin{eqnarray}
\ln R_Q=\frac{\ln \left(\tau _Q(Q)/\tau _Q(0)\right)}{\ln \left(\tau
  _Q(\bar{\varepsilon })/\tau _Q(0)\right)}. 
\label{rown:maintaus}
\end{eqnarray}
Thus using (\ref{rown:alphaQ}) and (\ref{rown:RQ}) we find
\begin{eqnarray}
\frac{\tau _Q(Q)}{\tau _Q(0)}=R_Q^{1/\alpha _Q}.
\label{rown:taQta0}
\end{eqnarray}
Equations~(\ref{rown:alphaQ}) and (\ref{rown:maintaus}) show the
decisive role of the relaxation time $\tau_Q(\varepsilon )$, but its
{\it ab initio\/} derivation is difficult.  Note that
Fig.~\ref{figure:Fig_3_7_8_9} and Table~\ref{table:tab_3_7_8_9} show a
data collapse for a given (fixed) value of a single control (aggregated)
variable $R_Q$.

\begin{table}
\centering
\caption{Values of exponent $\alpha _Q$ and quantity $\tau _Q(Q)$
  obtained from the fit of formula (\ref{rown:psiDtfin}) to the
  empirical data representing companies shown in
  Fig. \ref{figure:Fig_3_7_8_9} in plots (a), (b), (c) and (d) for
  $R_Q=$2, 5, 10, 30, 70.}
\begin{tabular}{|c||c|c|c|c|c|c|c|c|}
  \hline 
  & \multicolumn{2}{|c|}{Fig. \ref{figure:Fig_3_7_8_9}(a)} & \multicolumn{2}{|c|}{Fig. \ref{figure:Fig_3_7_8_9}(b)} & \multicolumn{2}{|c|}{Fig. \ref{figure:Fig_3_7_8_9}(c)} & \multicolumn{2}{|c|}{Fig. \ref{figure:Fig_3_7_8_9}(d)} \\
$R_Q$  & $\alpha _Q$ & $\tau _Q(Q)$ & $\alpha _Q$ & $\tau _Q(Q)$ & $\alpha _Q$ & $\tau _Q(Q)$ & $\alpha _Q$ & $\tau _Q(Q)$ \\ 
\hline  \hline
  2 & 1000 & 1.7699 & 1000 & 1.5436 & 1000 & 1.6129 & 1000 & 1.6129 \\
  \hline
  5 & 3.50 & 3.125 & 2.30 & 2.70 & 3.30 & 3.330 & 3.60 & 3.70 \\
  \hline 
	10 & 2.10 & 5.0 & 2.0 & 5.0 & 2.0 & 4.550 & 2.10 & 5.0 \\
  \hline
	30 & - & - & 1.050 & 5.560 & 1.0 & 5.0 & 1.10 & 5.260 \\
	\hline
	70 & - & - & 0.550 & 4.760 & 0.550 & 6.670 & 0.50 & 5.260 \\
  \hline
\end{tabular} 
\label{table:tab_3_7_8_9}
\end{table}

We analytically prove that $R_Q$ is the control variable that allows a
universal form of (\ref{rown:psiDtfin}) that depends solely on $R_Q$.
This variable was similarly used previously in connection with the
$q$-exponential \cite{LTB}.  Using this universal form requires that we
assume that the $B_Q$ in (\ref{rown:taue}) and (\ref{rown:alphaQ}))
depends on $Q$ in a power-law form, or the relevant scaling relation of
scaling variable $Q$,
\begin{eqnarray}
B_Q=B^{1/\eta }\frac{Q^\zeta }{\bar{\varepsilon }^{1+\zeta
}}=\frac{B^{1/\eta }}{\bar{\varepsilon }}\ln ^{\zeta /\eta }R_Q, 
\label{rown:BQxx}
\end{eqnarray}
where prefactor $B$ and exponent $\zeta $ are $Q$-independent positive
control parameters. Note that the second equality is a scaling relation
having $\ln R_Q$ as a scaling variable. Thus from (\ref{rown:alphaQ}),
(\ref{rown:RQ}), and (\ref{rown:BQxx}) we obtain the superscaling of the
scaling variable $\ln R_Q$ (or the scaling of scaling, i.e., the scaling
of the scaling exponent),
\begin{eqnarray}
\frac{1}{\alpha _Q}=B \ln ^{\zeta} R_Q,
\label{rown:alphaQxx}
\end{eqnarray}
which we further verify by examining data, e.g., for the IBM company,
which are typical of the empirical data used.

From (\ref{rown:taue}), (\ref{rown:RQ}), and (\ref{rown:taQta0}) we next
obtain the scaling relation of the scaling variable $\ln R_Q$,
\begin{eqnarray}
\frac{\tau _Q(Q)}{\tau _Q(0)}=\exp \left(\ln ^{1+\zeta }R_Q^{B^{
    1/(1+\zeta )}}\right)\Leftrightarrow   
\ln\left(\tau _Q(Q)/\tau _Q(0)\right)=B\ln ^{1+\zeta }R_Q.
\label{rown:taQta0xx}
\end{eqnarray}
Note that quantities $B_Q$, $\frac{1}{\alpha _Q}$, and $\frac{\tau
  _Q(Q)}{\tau _Q(0)}$ all depend on the single control variable $\ln
R_Q$. We will describe and discuss the $R_Q$-dependence of $\tau _Q(0)$
below, and will use the corresponding empirical data to confirm all the
$R_Q$-dependencies.

\subsection{Empirical verification of our formulas} 

\begin{figure}
\begin{center}
\includegraphics[width=145mm,angle=0,clip]{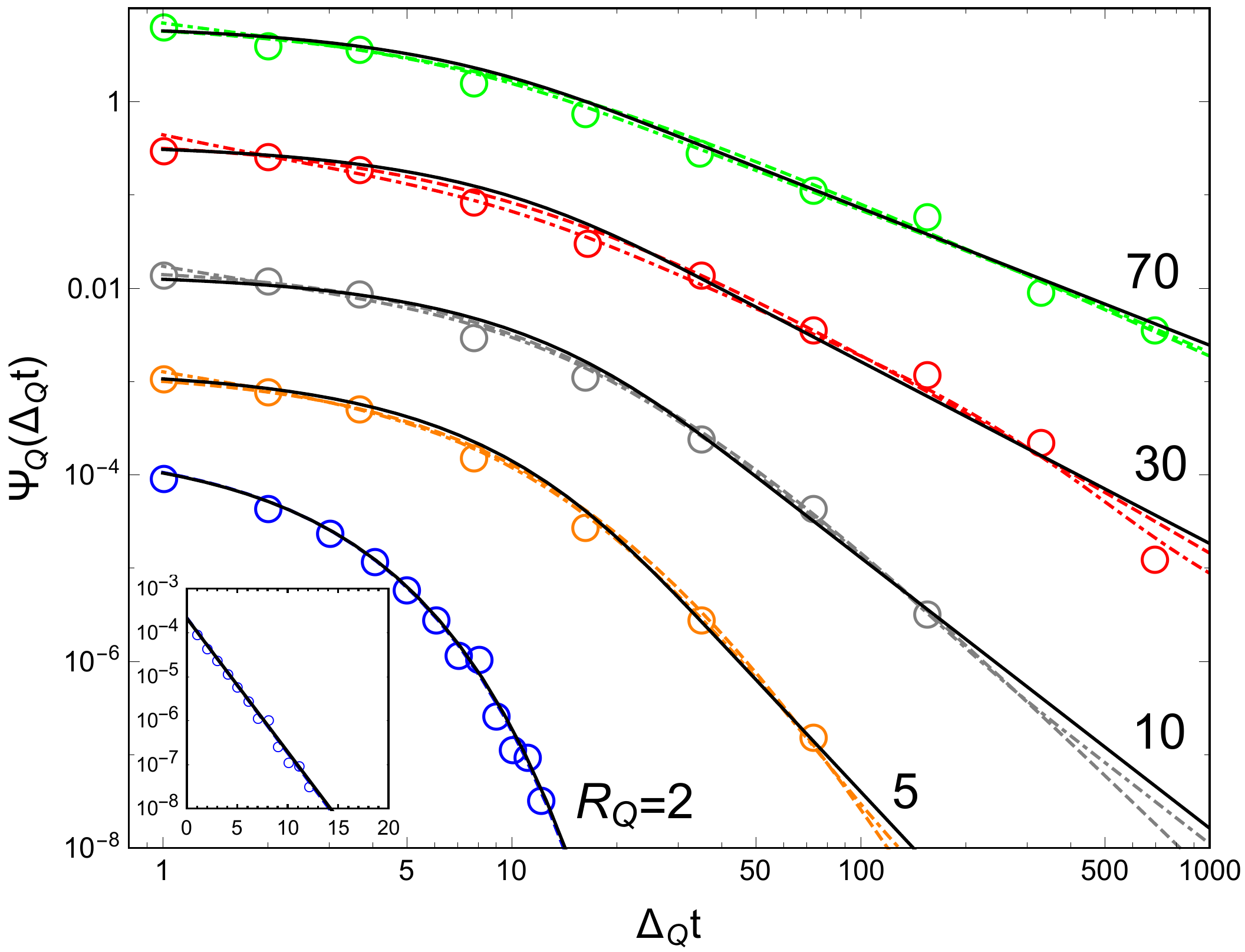}
\caption{The superstatistics, $\psi _Q(\Delta _Qt)$, vs. interevent
  time, $\Delta _Qt$, in log-log scale for the daily price returns of
  IBM (empty colored circles) in the period 1962-2010 for $R_Q$=2, 5,
  10, 30 and 70 (in units of days). Black solid curves are predictions
  of our formula (\ref{rown:psiDtfin}) while the dashed curves were
  given by q-exponential shown by eq. (3) in \cite{LTB}. The
  dashed-dotted curves are considered in Sec. \ref{section:disconc}. For
  $R_Q\geq 5$ the power-law relaxation of $\psi _Q(\Delta _Qt)$ is well
  seen for $\Delta _Qt>30$. The inset is the plot of $\psi _Q(\Delta
  _Qt)$ vs. $\Delta _Qt$ in the semi-logarithmic scale for $R_Q$=2 to
  clearly present the exponential form of the superstatistics. This
  exponential form was expected due to eq. (\ref{rown:psiDtinitexp}) as
  $\alpha _Q $ is very large in this case (see tab. \ref{table:tab2_0}).
  (Empirical data were drawn from \cite{LTB} for permission of EPL.)}
\label{figure:fig3a_EPL}
\end{center}
\end{figure}

Figure \ref{figure:fig3a_EPL} shows the agreement between the
predictions of (\ref{rown:psiDtfin}) and the empirical data for IBM for
$R_Q=2$, 5, 10, 30, and 70.

Table \ref{table:tab2_0} shows the fit of quantities $\alpha_Q$ and
$\tau_Q(Q)$.

\begin{table}
\centering
\caption{Values of exponent $\alpha _Q$ and quantity $\tau _Q(Q)$
  obtained directly from the fit of formula (\ref{rown:psiDtfin}) to the
  empirical data representing IBM company shown in
  Fig. \ref{figure:fig3a_EPL} for $R_Q$=2, 5, 10, 30, and 70.}
 \begin{tabular}{|c||c|c|c|}
  \hline 
 $R_Q$ & $Q$ & $\alpha _Q$ & $\tau _Q(Q)$ \\ 
\hline  \hline
  2 & 0.0050 & 1000 & 1.4286 \\
  \hline
  5 & 0.01389 & 3.0 & 3.330 \\
  \hline 
	10 & 0.02145 & 1.90 & 5.0 \\
	\hline 
	30 & 0.03442 & 0.950 & 4.550 \\
	\hline 
	70 & 0.04508 & 0.470 & 3.850 \\
  \hline
\end{tabular} 
\label{table:tab2_0}
\end{table}

\begin{figure}
\begin{center}
\includegraphics[width=110mm,angle=0,clip]{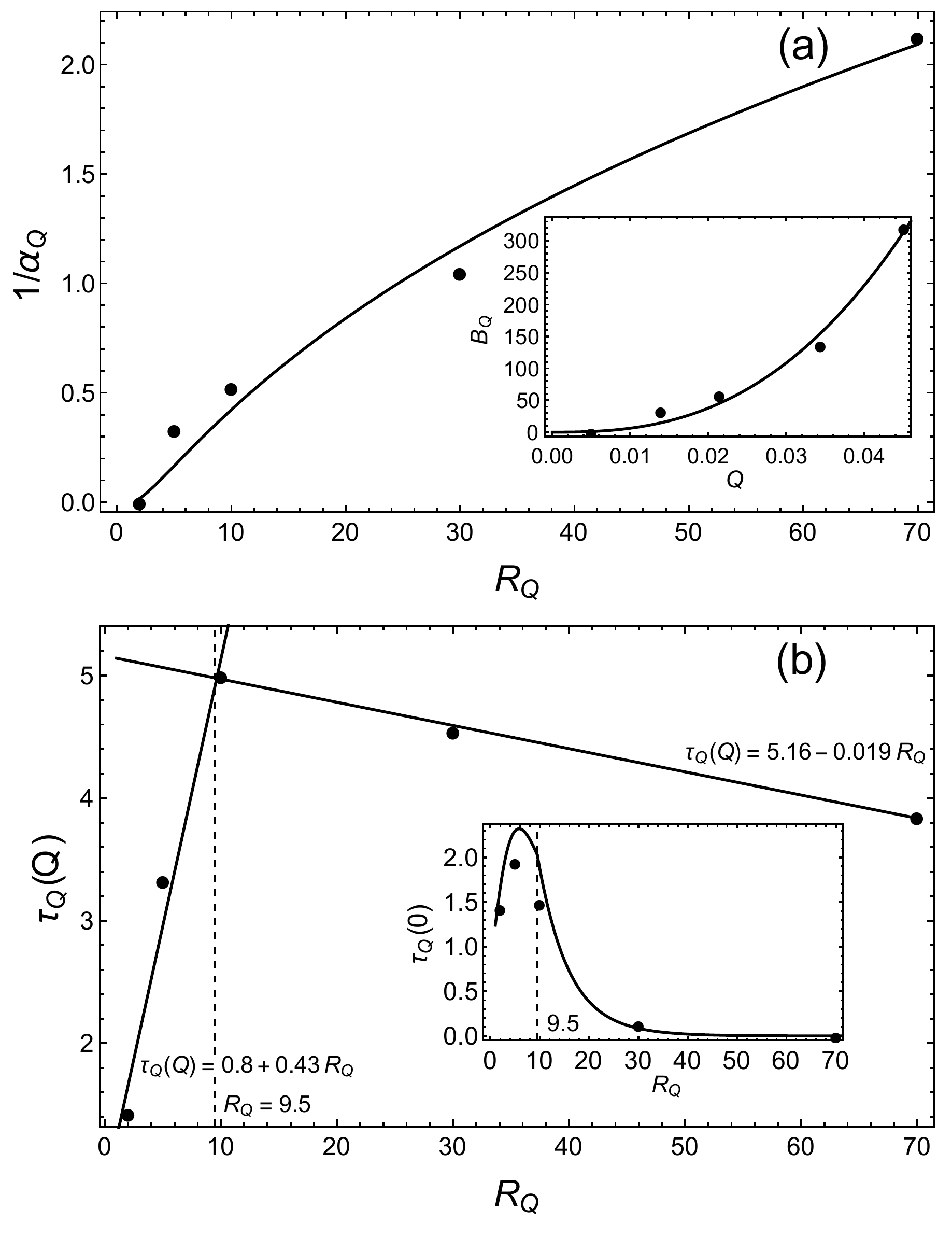}
\caption{Key dependence of quantities: (a) $1/\alpha _Q$ and (b)
  $\tau_Q(Q)$ vs. $R_Q$ obtained, for instance, for the IBM
  company. Black circles in main plots represent empirical data (shown
  in tab. \ref{table:tab2_0}), while solid curves are theoretical
  predictions. For (a) the solid curve was obtained by the fit of
  formula (\ref{rown:alphaQxx}) to empirical data, where the fit
  parameters $B$ and $\zeta $ were shown in tab. \ref{table:tabBQQ}. The
  indirect empirical data for the inset plot were found from
  eq. (\ref{rown:alphaQ}), where $\alpha _Q$ was taken from
  tab. \ref{table:tab2_0}, while $\eta $ and $\bar{\varepsilon }$ from
  tab. \ref{table:tab1} for the IBM company. The solid curve in this
  inset plot is the prediction of eq. (\ref{rown:BQxx}) for mentioned
  above parameters $B$ and $\zeta $.  For (b) the broken line or both
  solid straight lines are linear regressions (i.e. given by
  $\tau_Q(Q)=a_s\, R_Q+b_s,\; \mbox{where}\; s=L\; \mbox{for the lhs
    straight line}$ and $s=R\; \mbox{for the rhs straight line}$; here
  $\tau_0(0)=a_L+b_L=1.24$ as $R_{Q=0}=1$). Multiplicative and additive
  calibration parameters $a_s$ and $b_s$ defining both straight lines
  are shown in tab. \ref{table:theta_lines}. Hence, we have an
  additional interpretation of $\tau _Q(Q)$ as equal $R_Q$ up to some
  multiplicative and additive calibration parameters. The solid curve in
  the inset plot (i.e. $\tau _Q(0)$ vs. $R_Q$) was obtained from formula
  (\ref{rown:taQta0xx}), where $B$ and $\zeta$ comes from
  tab. \ref{table:tabBQQ}, while $\tau _Q(Q)$ was defined by above given
  straight lines.}
\label{figure:zbiorczy4tau}
\end{center}
\end{figure}

Figure \ref{figure:zbiorczy4tau}(a) shows the good fit of
(\ref{rown:alphaQxx}) (solid curve) to the data (black circles) also
found in the third column in Table~\ref{table:tab2_0}.

This fit allows us to determine the parameter $B$ and exponent $\zeta$
(see Table~\ref{table:tabBQQ}).

\begin{table}
\centering
\caption{Universal parameter $B$ and universal exponent $\zeta $,
  defining dependence of $B_Q$ vs. $Q$, obtained from the good fit of
  formula (\ref{rown:alphaQxx}) (solid curve) to the empirical data
  (black circles) shown in Fig. \ref{figure:zbiorczy4tau}a, for
  instance, for very representative IBM company.}
 \begin{tabular}{|c|c|}
  \hline 
 $B$ & $\zeta $ \\ 
\hline  \hline
  0.04798$\pm 0.0249$ & 2.6096$\pm 0.3478$ \\
  \hline 
\end{tabular} 
\label{table:tabBQQ}
\end{table}

The inset plot shows the good agreement between the (\ref{rown:BQxx})
prediction (solid curve) and the data (black circles). We prepared the
data by putting the first equality in (\ref{rown:alphaQ}) into the third
column of Table~\ref{table:tab2_1}, where the $\bar{\varepsilon}$ and
$\eta$ of IBM is supplied in Table~\ref{table:tab1}.

\begin{table}
\centering
\caption{Values of elementary quantities $B_Q$ and $\tau _Q(0)$ derived
  from key quantities $\alpha _Q$ and $\tau _Q(Q)$ found from the fit of
  formula (\ref{rown:psiDtfin}) to the empirical data representing IBM
  company shown in Fig. \ref{figure:fig3a_EPL} for $R_Q$=2, 5, 10, 30,
  and 70 for $\eta $=0.8246 and $\bar{\varepsilon }$=0.0078 given in
  tab. \ref{table:tab1}.  }
 \begin{tabular}{|c||c|c|c|}
  \hline 
 $R_Q$ & $Q$ & $B_Q$ & $\tau _Q(0)$ \\ 
\hline  \hline
 $2$ & $0.0050$ & $0.0295$ & $1.4286$ \\
  \hline 
  $5$ & $0.01389$ & $33.82951$ & $1.94745$ \\
  \hline 
	$10$ & $0.02145$ & $58.86516$ & $1.48793$ \\
	\hline 
	$30$ & $0.03442$ & $136.43324$ & $1.26807\times 10^{-1}$ \\
	\hline 
	$70$ & $0.04508$ & $320.29882$ & $4.56615\times 10^{-4}\approx 0$ \\
  \hline
\end{tabular} 
\label{table:tab2_1}
\end{table}

Figure \ref{figure:zbiorczy4tau}(b) shows a plot of $\tau _Q(Q)$
vs. $Q$, where $\tau _Q(Q)$ comes from the fourth column of
Table~\ref{table:tab2_0}.  The plot consists of a broken straight line
or two crossing straight lines. Table~\ref{table:theta_lines} shows the
parameters of linear regressions $a_s$ and $b_s$, with $s=L,R$, that
define the dependence of both straight lines on $R_Q$. The inset plot
uses (\ref{rown:taQta0}) and (\ref{rown:alphaQxx}) to calculate (i) the
data points (black circles) with $\tau_Q(Q)$ supplied by
Table~\ref{table:tab2_0}, and (ii) the solid curve, using the analytical
form of $\tau_Q(Q)$.

\begin{table}
\centering
\caption{Parameters of linear regressions $a_s$ and $b_s$, $s=L,R$,
  defining dependence of both straight lines on $R_Q$, presented in
  Fig. \ref{figure:zbiorczy4tau}b for the IBM company.}
 \begin{tabular}{|c|c|c|}
  \hline 
 Parameters & $L$ & $R$ \\ 
\hline  \hline
  $a_s$ & 0.430 & -0.019 \\
  \hline 
  $b_s$ & 0.80 & 5.160 \\
  \hline
\end{tabular} 
\label{table:theta_lines}
\end{table}
  
Thus by analytically and empirically proving the $R_Q$-dependence of the
superstatistics $\psi_Q(\Delta_Q t)$ we explain the empirical data
collapse shown in Fig.~\ref{figure:Fig_3_7_8_9}.
 
\section{Discussion and concluding remarks}\label{section:disconc}

We find an explicit closed form of the threshold interevent time
superstatistics (\ref{rown:psiDtfin}) that is valid for excessive
losses, is the foundation of the continuous time random walk (CTRW)
formalism, and that is useful in the study of a double action market
(see \cite{GK} and refs. therein). These superstatistics are more
credible than the $q$-exponential distribution that is applied {\it
  ad hoc\/} in this context in Ref.~\cite{LTB,LB}, and they agree with
the key empirical relation between the mean interevent time $R_Q$ and
the threshold $Q$ (see Fig.~\ref{figure:fig.2}).

We model the empirical data collapse (cf. Fig.~\ref{figure:Fig_3_7_8_9})
using superstatistics as a function of a single aggregated variable
$R_Q$ and obtain, for example, the scaling shape exponent $1/\alpha_Q$
as a power-law function of $\ln R_Q$ and the superscaling form of
(\ref{rown:alphaQxx}) that is dependent upon universal exponent $\zeta$
and prefactor $B$.

\begin{figure}
\begin{center}
\includegraphics[width=155mm,angle=0,clip]{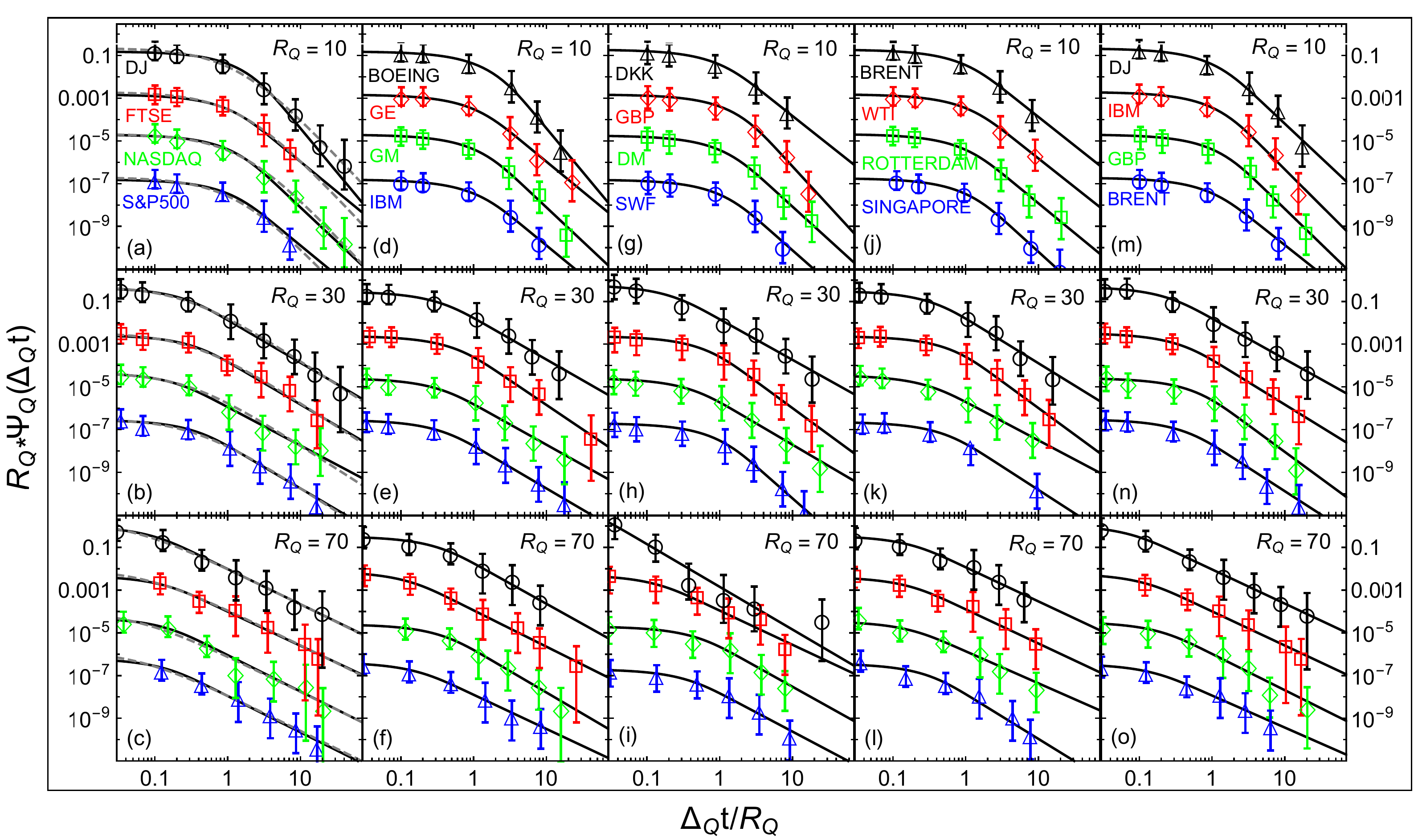}
\caption{Statistics of interevent times between arithmetic profit
  returns of daily closing for various markets (from stock exchange and
  forex to resource market) and time periods. All empirical data
  (discrete marks with bars) were taken from ref. \cite{BB1}. Solid
  curves are predictions of our formula (\ref{rown:psiDtfin}) as it can
  be applied both for losses and profits. Dashed curves shown, for
  instance, in plots (a), (b), and (c) are fitted by q-exponential
  (remaining twelve plots are very similarly fitted therefore, the fits
  are not visualized herein). However, the possible empirical data
  collapse would be incredible in this case because errors of empirical
  data points are too large. (Empirical data were drawn from \cite{BB1}
  for permission of PRE.)}
\label{figure:FitFig7a-11cPRE}
\end{center}
\end{figure}

Note that (\ref{rown:psiDtfin}) accurately describes the empirical
statistics of excessive profits. Here $Q$ defines the threshold for
excessive profits instead of excessive losses (see the plots in
Fig.~\ref{figure:FitFig7a-11cPRE}). We thus can use the same
superstatistics to demonstrate the functional but not literal symmetry
between excessive losses and profits. The symmetry is not literal
because different control parameters $B$ and $\zeta$ and driving
parameters $a_s$ and $b_s$ ($s=L,R$) are used.  Because of large
statistical errors in the empirical data, we cannot empirically verify
the universality of excessive profit behavior.  For example, for
$R_Q=10$ exponent $1.70\leq \alpha _Q\leq 3.10$ and $0.10\leq
\tau_Q(Q)\leq 0.25$, for $R_Q=30$ we have $0.90\leq \alpha _Q\leq 1.50$
and $0.12\leq \tau_Q(Q)\leq 0.35$, and finally for $R_Q=70$ we have
$0.60\leq \alpha _Q\leq 1.10$ and $0.08\leq \tau_Q(Q)\leq 0.36$, which
exhibit ranges that are too extended.

Note that if we substitute $\tau _Q^{\prime }(\varepsilon)$ given by
(\ref{rown:tauen})---the case opposite to that defined by
(\ref{rown:taue})---into (\ref{rown:psiDte}), and use a derivation
analogous to the one that produced (\ref{rown:psiDtfin}), we obtain a
result complementary to (\ref{rown:psiDtfin}), i.e.,
\begin{eqnarray}
\psi _Q^{\prime }(\Delta _Q t)&=&\frac{1}{\tau _Q^{\prime
  }(Q)}\frac{\alpha _Q}{(\Delta _Q t/\tau _Q^{\prime }(Q))^{1+\alpha
    _Q}} \nonumber \\ &\times &{\bf \gamma }_{Euler}^{\prime }(1+\alpha
_Q,\Delta _Q t/\tau _Q^{\prime }(Q)),
\label{rown:psiDtfinn}
\end{eqnarray}
where 
\begin{eqnarray}
&&{\bf \gamma }_{Euler}^{\prime }(1+\alpha _Q,\Delta _Q t/\tau
  _Q^{\prime }(Q))=\int_{\Delta _Q t/\tau _Q^{\prime }(Q)}^{\infty }
  y^{\alpha _Q}\exp(-y)dy\nonumber \\ &=&{\bf \Gamma }_{Euler}(1+\alpha
  _Q)-{\bf \gamma }_{Euler}(1+\alpha _Q,\Delta _Q t/\tau _Q^{\prime
  }(Q)),
\label{rown:gameuln}
\end{eqnarray}
is the upper gamma function, which for $\Delta _Q t/\tau
_Q^{\prime}(Q)\gg 1$ truncates the power-law in (\ref{rown:psiDtfinn}).
In the opposite case of $\Delta _Q t/\tau_Q^{\prime }(Q)\ll 1$ we obtain
\begin{eqnarray}
\psi _Q^{\prime }(\Delta _Q t)\approx \frac{1}{\tau _Q^{\prime
  }(Q)}\frac{\alpha _Q}{(\Delta _Q t/\tau _Q^{\prime }(Q))^{1+\alpha
    _Q}}{\bf \Gamma }_{Euler}(1+\alpha _Q),
\label{rown:psiDtasymptn}
\end{eqnarray}
which is only formally identical to (\ref{rown:psiDtasympt}).
Figure~\ref{figure:fig3a_EPL} shows the best predictions of the weighted
sum of superstatistics (dashed dotted curves) given by
(\ref{rown:psiDtfin}) and (\ref{rown:psiDtfinn}).  Note that this
continues to agree with the corresponding empirical data for IBM.

We use (\ref{rown:rho}), (\ref{rown:taue}), and (\ref{rown:alphaQ}) to
obtain the moment $\langle (\Delta _Q t)^m \rangle ,\, m=0,1,2,\ldots$
in an explicit closed form,
\begin{eqnarray}
\langle (\Delta _Q t)^m \rangle &=& \frac{\int_0^{\infty }(\Delta _Q
  t)^m\, \psi _Q(\Delta _Q t)d(\Delta _Q t)}{\int _Q^{\infty
  }D(\varepsilon )d\varepsilon} \nonumber \\ &=&\frac{m!\int _Q^{\infty
  }[\tau _Q(\varepsilon )]^mD(\varepsilon )d\varepsilon }{\int
  _Q^{\infty }D(\varepsilon )d\varepsilon } \nonumber
\\ &=&R_Q(\tau
_Q(Q))^m\frac{m!}{1-m/\alpha _Q},\;\; m=0,1,2,\ldots ,
\label{rown:meanDt}
\end{eqnarray}
where the first equality gives the definition\footnote{Here we consider
  only integer non-negative moments.}. Note that the moments of
arbitrary order, as well as $\alpha _Q$ and $\tau_Q(Q)$, depend solely
on $R_Q$. Note also that $\langle (\Delta _Q t)^m \rangle $ is finite
only for $\alpha _Q>m$, and that otherwise it diverges. This is in
contrast to the behavior of $R_Q$, which, because of its quantile (not
momentum) origin, is always finite, e.g., for IBM $\langle \Delta _Q t
\rangle $ is finite only for $R_Q\leq 10$ (see
Table~\ref{table:tab2_0}). We thus have two radically different cases,
finite interevent time and infinite interevent time $\langle\Delta _Q t
\rangle$, about which there is much in the literature (see, e.g.,
\cite{ELIB,RKFS,RKMR1,RKMR2} and refs. therein). Infinite interevent
time is particularly interesting when it takes into consideration
ergodicity breaking \cite{BelBar1,BelBar2}.

Note that using our microscopic model to simulate agent behavior
\cite{DGK,DGKarXiv} gives results very close to those predicted by
(\ref{rown:psiDtfin}). An approach using agent-based modeling in this
context was recently explored by other authors \cite{Gont}.

It is our hope that this work will constitute a strong contribution to
the research effort searching for universal properties in market
behavior.


\bibliographystyle{Science}





\begin{scilastnote}
\item{\bf Acknowledgments} 
 Two of us (M.J. and T.G.) are grateful to
  the Foundation for Polish Science for financial support.
 The work of
  H.E.S.  was supported by NSF Grant CMMI 1125290, DTRA Grant
  HDTRA1-14-1-0017 and ONR Grant N00014-14-1-0738.  One of us (R.K.) is
  grateful for inspiring discussions with Shlomo Havlin (during The
  Third 
 Nikkei Econophysics Symposium, Tokyo 2004), Armin Bunde
  (during his visit at Faculty of Physics 
 University of Warsaw in
  2011), and Constantino Tsallis (during the SMSEC2014 in Kobe).
\end{scilastnote}

\end{document}